\documentclass[twocolumn,prl,superscriptaddress,noshowpacs,epsf]{revtex4}

\usepackage{graphicx}

\begin{document}
\title{Topological states and braiding statistics using quantum circuits}
\date{\today}
\author{J. Q. You}
\affiliation{Department of Physics and Surface Physics Laboratory
(National Key Laboratory), Fudan University, Shanghai 200433, China}
\affiliation{Advanced Study Institute, The Institute of Physical and
Chemical Research (RIKEN), Wako-shi 351-0198, Japan}
\author{Xiao-Feng Shi}
\affiliation{Department of Physics and Surface Physics Laboratory
(National Key Laboratory), Fudan University, Shanghai 200433, China}
\affiliation{Advanced Study Institute, The Institute of Physical and
Chemical Research (RIKEN), Wako-shi 351-0198, Japan}
%
\author{Franco Nori}
\affiliation{Advanced Study Institute, The Institute of Physical and
Chemical Research (RIKEN), Wako-shi 351-0198, Japan}
\affiliation{Center for
Theoretical Physics, Physics Department, Center for the Study of
Complex Systems, University of Michigan, Ann Arbor, MI 48109-1040,
USA}

\begin{abstract}
Using superconducting quantum circuits, we propose an approach to
construct a Kitaev lattice, i.e., an anisotropic spin model on a
honeycomb lattice with three types of nearest-neighbor interactions.
We study two particular cases to demonstrate topological states
(i.e., the vortex and bond states) and show how the braiding
statistics can be revealed. Our approach provides an experimentally
realizable many-body system for demonstrating exotic properties of
topological phases.
\end{abstract}
\maketitle

Topological quantum systems are currently attracting considerable
interest because of their fundamental importance and potential
applications in quantum computing~\cite{RMP}. They exhibit
topological phases of matter that are insensitive to local
perturbations. This exotic property makes them appealing for
fault-tolerant quantum computing.

Topological quantum computing can be implemented by storing and
processing quantum information with anyons~\cite{RMP}, which are
neither bosons nor fermions, but obey anyonic braiding statistics.
To achieve this, it is a centrally important issue to experimentally
realize a topological quantum system. Recently, topological quantum
computing using non-Abelian anyons in a fractional quantum Hall
system was proposed~\cite{Sarma05}; however, it has not been
experimentally verified if the observed fractional quantum Hall
states are the desired topological states for quantum computing.
Indeed, up to now, anyons have never been directly observed in
experiments. Thus, in order to guarantee anyonic topological phases
in a many-body quantum system, one could design artificial, but
exactly solvable lattice models that have such desired topological
phases. The most important example is the Kitaev model on a
honeycomb lattice~\cite{Kitaev}, which is an anisotropic spin model
with three types of nearest-neighbor interactions. Most strikingly,
depending on the bond parameters, this model~\cite{Kitaev} supports
both Abelian and non-Abelian anyons. Therefore, if this topological
model could be realized in experiments, it would provide exciting
opportunities for experimentally demonstrating anyons and
implementing topological quantum computing.

Proposals were made for implementing the Kitaev model using atomic
optical lattices~\cite{Duan, Zoller}. Also, it was
shown~\cite{Sarma07} that optical lattices can be used to
demonstrate the anyonic braiding statistics. However, its
implementation using an optical lattice requires extremely low
temperatures~\cite{Duan} that are beyond current experimental
capabilities. Moreover, to demonstrate anyonic braiding, the optical
lattice needs to perform rotations on single atoms at selected sites
using external laser beams. This is difficult because the atomic
spacing is of the order of laser wavelength and the diffraction
limit takes effect.

Here we propose an approach to construct a Kitaev lattice using
superconducting quantum circuits~\cite{YN05}. A charge qubit is
placed at each site and the three types of nearest-neighbor
interactions are achieved using different circuit elements. This
circuit can demonstrate topological states (i.e., the vortex and
bond states) and show how the braiding statistics can be revealed.
Here, the quantum circuits behave like artificial spins~\cite{YN05}
and the Kitaev model can be constructed using the experimentally
available technologies for superconducting qubits. Moreover, these
charge qubits are tunable via external fields, making the required
single-qubit rotations implementable for demonstrating the braiding
statistics of the topological states. Our proposal provides an
experimentally realizable many-body system for demonstrating exotic
properties of the topological phases.

\begin{figure}
\includegraphics[width=3.1in,  
bbllx=100,bblly=213,bburx=467,bbury=725]{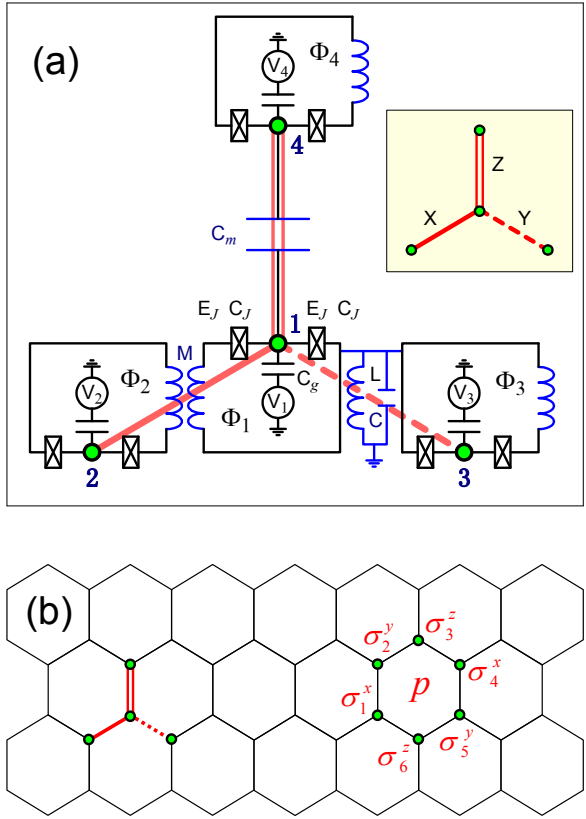} \caption{(Color
online) (a)~Schematic diagram of the basic building block of a
Kitaev lattice, consisting of four superconducting charge qubits
(labelled 1 to 4): (i)~The qubits 1 and 2 are inductively coupled
via a mutual inductance $M$; (ii)~the qubits 1 and 3 are coupled via
an $LC$ oscillator; and (iii)~the qubits 1 and 4 are capacitively
coupled via a mutual capacitance $C_m$. Inset:~These three types of
inter-qubit couplings are denoted as $X$, and $Y$ and $Z$ bonds.
Here each charge qubit consists of a Cooper-pair box (green dot)
which is linked to a superconducting ring, via two identical
Josephson junctions (each with coupling energy $E_J$ and capacitance
$C_J$), to form a SQUID loop. Also, each qubit is controlled by both
a voltage $V_i$ (applied to the qubit via the gate capacitance
$C_g$) and a magnetic flux $\Phi_i$ (piercing the SQUID loop). (b) A
subset of the Kitaev lattice (honeycomb lattice) constructed by
repeating the building block in (a), where a charge qubit is placed
at each site. Also, the plaquette operator
$W_p=\sigma_1^x\sigma_2^y\sigma_3^z\sigma_4^x\sigma_5^y\sigma_6^z$
is shown for a given plaquette (hexagon) $p$.} \label{fig1}
\end{figure}

{\it Kitaev model.}{\bf---}We use superconducting quantum circuits
to construct an artificial Kitaev lattice (see Fig.~\ref{fig1}). The
building block of the lattice consists of four charge qubits
[Fig.~\ref{fig1}(a)], each placed at a lattice site of a hexagonal
lattice [Fig.~\ref{fig1}(b)]. Here, each charge qubit is a
Cooper-pair box connected to a superconducting ring by two identical
Josephson junctions. Each qubit is controlled by both the magnetic
flux $\Phi_i$ piercing the SQUID loop and the voltage $V_i$ applied
via the gate capacitance $C_g$. Moreover, as shown in
Fig.~\ref{fig1}(a), the nearest-neighboring charge qubits in the
three different link directions are coupled by different circuit
elements:

(i)~{\it $X$-type bond}. The two qubits [denoted by 1 and 2 in
Fig.~\ref{fig1}(a)] in the $x$-link direction are coupled by a
mutual inductance $M$. This inductive coupling is given
by~\cite{YTN} $K_x(1,2)=MI_1I_2$, where the circulating supercurrent
$I_i$ in the SQUID loop of the $i$th charge qubit is
$I_i=I_c\sin(\pi\Phi_i/\Phi_0)\cos\varphi_i$, with $I_c=2\pi
E_J/\Phi_0$, $\Phi_0=h/2e$ (the flux quantum) and $\varphi_i$ being
the average phase drop across the two Josephson junctions.

(ii)~{\it $Y$-type bond}. The two qubits 1 and 3 in the $y$-link
direction are coupled via an $LC$ oscillator~\cite{Schon} and the
inter-qubit coupling is $K_y(1,3)=-4\xi
E_{J1}(\Phi_1)E_{J3}(\Phi_3)\sin\varphi_1\sin\varphi_3$, where
$E_{Ji}(\Phi_i)=2E_J\cos(\pi\Phi_i/\Phi_0)$, and $\xi=L[\pi
C_{\Sigma}(C_g+C_m)/\Lambda\Phi_0]^2$, with
$C_{\Sigma}=2C_J+C_g+C_m$, and $\Lambda=C_{\Sigma}^2-C_m$. Note that
$\xi\propto (C_g+C_m)^2$ in the presence of the mutual capacitance
$C_m$ that connects qubit 1 (3) with its nearest-neighboring qubit
in the vertical ($z$-link) direction. Usually, $C_m\gg C_g$, so the
coupling between qubits 1 and 3 is now much increased, compared to
the usual case without the mutual inductance~\cite{Schon}, where
$\xi\propto C_g^2$.

(iii)~{\it $Z$-type bond}. The two qubits 1 and 4 in the $z$-link
direction are coupled via the mutual capacitance
$C_m$~\cite{NEC,Bruder}: $K_z(1,4)=E_m(n_1-n_{g1})(n_4-n_{g4})$,
where $E_m=4e^2C_m/\Lambda$, $n_{gi}=C_gV_i/2e$, and
$n_i=-i\partial/\partial\varphi_i$ is a number operator of the
Cooper pairs in the $i$th box. The mutual capacitance $C_m$ modifies
the charging energy $2e^2/C_{\Sigma}$ of an isolated charge qubit to
become $E_c=2e^2C_\Sigma/\Lambda$.

When the electrostatic and the Josephson coupling energies of each
charge qubit are included, the Hamiltonian of the lattice is
$H=\sum_i[E_c(n_i-n_{gi})^2 - E_{Ji}(\Phi_i)\cos\varphi_i]
+\sum_{\rm x-link}K_x(j,k)+\sum_{\rm y-link}K_y(j,k) +\sum_{\rm
z-link}K_z(j,k)$.
For charge qubits, $E_c\gg E_J$. When the gate voltage $V_i$ is near
the optimal point $e/C_g$, i.e., $n_{gi}\sim \frac{1}{2}$, only two
charge states $|0\rangle_i$ and $|1\rangle_i$, corresponding to zero
and one extra Cooper pairs in the box, are important for each qubit.
In the spin-$\frac{1}{2}$ representation, based on the charge states
$|0\rangle_i\equiv|\!\!\uparrow\rangle_i$ and
$|1\rangle_i\equiv|\!\!\downarrow\rangle_i$, one has
$n_i=\frac{1}{2}(1-\sigma_i^z)$,
$\cos\varphi_i=\frac{1}{2}\sigma_i^x$, and
$\sin\varphi_i=-\frac{1}{2}\sigma_i^y$. Here we consider the simple
case with $n_{gi}=n_g$ (i.e., $V_i=V_g$) and $\Phi_i=\Phi_e$ for all
qubits. The Hamiltonian of the system is then reduced to
\begin{eqnarray}
H\!&\!=\!&\!J_x\sum_{\rm x-link}\sigma_j^x\sigma_k^x +J_y\sum_{\rm
y-link}\sigma_j^y\sigma_k^y +J_z\sum_{\rm
z-link}\sigma_j^z\sigma_k^z\nonumber\\
&&\!+\sum_i(h_z\sigma_i^z+h_x\sigma_i^x). \label{model}
\end{eqnarray}
This is the Kitaev model on a honeycomb lattice, in the presence of
a ``magnetic field" with $z$- and $x$-components. Here
$h_z=(E_c+\frac{1}{2}E_m)(n_{g}-\frac{1}{2})$, and
$h_x=-\frac{1}{2}E_{J}(\Phi_e)$, with
$E_{J}(\Phi_e)=2E_J\cos(\pi\Phi_e/\Phi_0)$. The bond parameters are
$J_x=\frac{1}{4}MI_c^2\sin^2(\pi \Phi_e/\Phi_0)\geq 0$, $J_y=-\xi
[E_{J}(\Phi_e)]^2\leq 0$, and $J_z=\frac{1}{4}E_m>0$. The magnetic
field can be used to achieve single-qubit rotations for
demonstrating anyonic braiding statistics.

{\it Topological excitations.}{\bf---}Below we focus on two
particular cases to show the properties of topological excitations
in the Kitaev model (\ref{model}):

(i)~{\it Kitaev lattice in a weak magnetic field}. We first consider
the case with $|h_z|,|h_x|\ll J_x,|J_y|,J_z$, and choose
$V=\sum_i(h_z\sigma_i^z+h_x\sigma_i^x)$ as the perturbation. Using
perturbation theory in the Green function formalism~\cite{Kitaev},
one can construct an effective Hamiltonian $H_{\rm eff}$ acting on
the vortex-free sector: $H_{\rm
eff}=-(2h_z^2/\Delta\varepsilon_z)\sum_{\rm
z-link}\sigma_j^z\sigma_k^z-(2h_x^2/\Delta\varepsilon_x)\sum_{\rm
x-link}\sigma_j^x\sigma_k^x$, where $\varepsilon_{z(x)}$ is the
excitation energy of the state $\sigma^{z(x)}|g\rangle$, and
$|g\rangle$ is the eigenstate of all plaquette operators $W_p$ [see
Fig.~\ref{fig1}(b)] corresponding to the eigenvalue $w_p=1$.
Here the effective Hamiltonian $H_{\rm eff}$ is only contributed by
the second-order term because both the first- and third-order terms
are zero. With the zeroth-order term (unperturbed Hamiltonian)
included, the Hamiltonian of the system can be equivalently written
as $H=J'_x\sum_{\rm x-link}\sigma_j^x\sigma_k^x +J_y\sum_{\rm
y-link}\sigma_j^y\sigma_k^y +J'_z\sum_{\rm
z-link}\sigma_j^z\sigma_k^z$, where the bond parameters $J_z$ and
$J_x$ are renormalized to $J'_z=J_z-2h_z^2/\Delta\varepsilon_z$, and
$J'_x=J_x-2h_x^2/\Delta\varepsilon_x$.

\begin{figure}
\includegraphics[width=3.3in,  
bbllx=133,bblly=215,bburx=494,bbury=753]{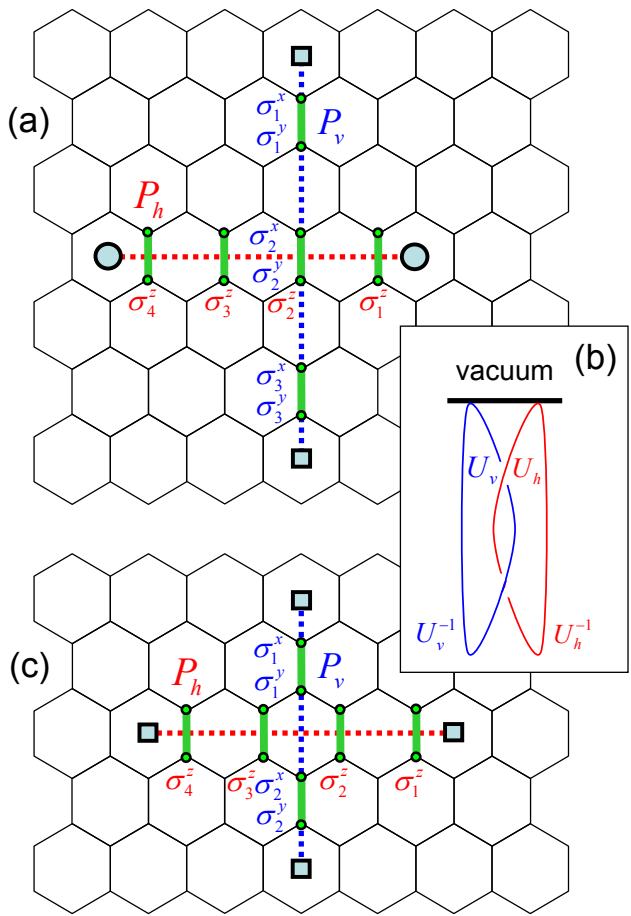} \caption{(Color
online) Schematic diagram of the procedures for braiding topological
excitations. (a)~The operations $U_h$ and $U_v$ for creating
different types of topological excitations, which are achieved by
successively applying spin-pair operators at $z$-bonds along the
horizontal ($P_h$) and vertical ($P_v$) paths. (b)~A combined
operation $U_h^{-1}U_v^{-1}U_hU_v$ for both braiding the topological
excitations created in (a) and fusing them to the vacuum. (c)~The
operations $U_h$ and $U_v$ for creating the same types of
topological excitations, which are achieved by successively applying
spin-pair operators at $z$-bonds along $P_h$ and $P_v$.
}\label{fig2}
\end{figure}

To show the topological excitations, we focus on the Abelian
case~\cite{Kitaev} with $J'_z\gg J'_x, |J_y|$. The dominant part of
$H$, i.e., $H_0=J'_z\sum_{\rm z-link}\sigma_j^z\sigma_k^z$, favors a
highly degenerate ground state with each pair of spins in each
z-link aligned opposite to each other
($|\!\!\uparrow\downarrow\rangle$ or
$|\!\!\downarrow\uparrow\rangle$). While preserving the ground-state
subspace, the effective Hamiltonian can now be written, up to fourth
order, as
\begin{equation}
H_{\rm eff}=-J_{\rm eff}\sum_p W_p, \label{vortex}
\end{equation}
where $J_{\rm eff}={J'}_x^2J_y^2/16{J'}_z^3$. Clearly, the ground
state of the system preserves $w_p=1$, i.e.,
$W_p|g\rangle=|g\rangle$, for all plaquettes $p$. A pair of vortex
excitations are created when the system changes with
$w_p=1\rightarrow-1$ for two neighboring plaquettes. This can be
achieved by acting a spin-pair operator on the ground state
$|g\rangle$:
$|\widetilde{Z}_i\rangle=\tilde{\sigma}_i^z|g\rangle$, and
$|\widetilde{Y}_i\rangle=\tilde{\sigma}_i^y|g\rangle$,
with $\tilde{\sigma}_i^z\equiv\sigma_{i}^zI_{i}$ and
$\tilde{\sigma}_i^y\equiv\sigma_{i}^y\sigma_{i}^x$. Here the two
operators $\sigma^z_i$ ($\sigma^y_i$) and $I_i$ ($\sigma^x_i$) act
on the ground state $|g\rangle$ at the bottom and top sites of the
$i$th z-link, respectively. This pair of vortex excitations
are Abelian anyons~\cite{Kitaev,Sarma07}, which have an energy gap
$\Delta\varepsilon=4J_{\rm eff}$ above the ground state.

(ii)~{\it Kitaev lattice with dominant $z$-bonds}. We choose
$h_z=0$, which corresponds to the case with each charge qubit
working at the optimal point $n_g=\frac{1}{2}$. Moreover, we choose
suitable circuit parameters
to have $J_z\gg |h_x|, J_x, |J_y|$, with $|h_x|$ comparable to $J_x,
|J_y|$. Here we also use perturbation theory in the Green function
formalism to derive the effective Hamiltonian. Up to the
second-order term, the effective Hamiltonian can be written as
\begin{equation}
H_{\rm eff}=-K_{\rm eff}\sum_{\rm z-link}\sigma_j^x\sigma_k^x,
\label{bond}
\end{equation}
with $K_{\rm eff}=h_x^2/J_z$. The spin-pair operator
$\sigma_j^x\sigma_k^x$ at a $z$-bond is commutative with the
unperturbed zero-order Hamiltonian $H_0=J_z\sum_{\rm
z-link}\sigma_j^z\sigma_k^z$, but anticommutative to the four
plaquette operators $W_p$ connected to this $z$-bond. Similar to
$W_p$, the pair operator $\sigma_j^x\sigma_k^x$ also has two
eigenvalues $p_z=\pm 1$. Now the ground state $|g\rangle$ of the
system preserves $p_z=1$, i.e.,
$\sigma_j^x\sigma_k^x\,|g\rangle=|g\rangle$, for all $z$-bonds. When
the pair operators $\tilde{\sigma}_i^z$ and $\tilde{\sigma}_i^y$ are
separately applied to the ground state at the $i$th $z$-bond, the
excited states
$|\widetilde{Z}_i\rangle=\tilde{\sigma}_i^z|g\rangle$, and
$|\widetilde{Y}_i\rangle=\tilde{\sigma}_i^y|g\rangle$ are two types
of bond states; each corresponding to the change $p_z=1\rightarrow
-1$ at the $i$th $z$-bond and having an energy gap $2K_{\rm eff}$
above the ground state.

{\it The braiding of topological excitations.}{\bf---}As shown in
\cite{Sarma07}, an $e$-type vortex looping around another $e$ vortex
does not produce a sign change to the wave function, while an $e$
vortex looping around an $m$-type vortex does. This indicates
anyonic statistics between the $e$ and $m$ vortex states. Here we
show an alternative procedure for braiding a topological excitation
with another, which applies to both the vortex and bond states.

First, successively apply spin-pair operations
$\tilde{\sigma}_i^y=\sigma_{i}^y\sigma_{i}^x$ to the ground state
$|g\rangle$ at three $z$-bonds in the vertical path $P_v$ [see
Fig.~\ref{fig2}(a)]. In the case of Hamiltonian (\ref{vortex}), this
creates a pair of $e$ vortex states and move a vortex along the
vertical path $P_v$, while a bond state is generated in the case of
Hamiltonian (\ref{bond}). Then, we successively apply the operations
$\tilde{\sigma}_i^z=\sigma_{i}^zI_{i}$ at four $z$-bonds along the
horizontal path $P_h$ [see also Fig.~\ref{fig2}(a)]. After these
operations, the state of the system is $U_hU_v|g\rangle$, where
$U_h=\tilde{\sigma}_4^z\tilde{\sigma}_3^z\tilde{\sigma}_2^z\tilde{\sigma}_1^z$,
and $U_v=\tilde{\sigma}_3^y\tilde{\sigma}_2^y\tilde{\sigma}_1^y$
correspond to the operations along the horizontal and vertical
paths. Furthermore, successively apply $U_v^{-1}$ and $U_h^{-1}$ to
the system, so as to fuse~\cite{RMP,Kitaev} the excitations to the
vacuum (i.e., the ground state) [see Fig.~\ref{fig2}(b)]. Now, the
final state of the system is
$|\Psi_f\rangle=U_h^{-1}U_v^{-1}U_hU_v|g\rangle$. Because the paths
$P_v$ and $P_h$ intersect at a lattice point, where
$\tilde{\sigma}^y_2$ and $\tilde{\sigma}^z_2$ anticommute, one has
$U_hU_v=-U_vU_h$. Therefore, the final state becomes
$|\Psi_f\rangle=-|g\rangle$.
In contrast, for two topological excitations of the same type, when
similar operations are applied, the paths $P_v$ and $P_h$ do not
intersect at a lattice point [see Fig.~\ref{fig2}(c)]. Thus,
$U_hU_v=U_vU_h$, and $|\Psi_f\rangle=|g\rangle$, yielding no sign
change to the ground-state wave function.

The braiding statistics of topological excitations can be revealed
by means of Ramsey-type interference~\cite{Sarma07,Pachos}. To
achieve this, we keep the same $U_v$ as above, but use
$U_h=(\tilde{\sigma}^z_4)^{\frac{1}{2}}(\tilde{\sigma}^z_3)^{\frac{1}{2}}
(\tilde{\sigma}^z_2)^{\frac{1}{2}}(\tilde{\sigma}^z_1)^{\frac{1}{2}}$,
where
$(\tilde{\sigma}^z_i)^{\frac{1}{2}}\equiv(\sigma^z_i)^{\frac{1}{2}}I_i$,
i.e., each $\sigma^z_i$ is replaced by half of the rotation. In the
braiding case shown in Fig.~\ref{fig2}(a),
$\tilde{\sigma}^y_2(\tilde{\sigma}^z_2)^{\frac{1}{2}}
=i(\tilde{\sigma}^z_2)^{-\frac{1}{2}}\tilde{\sigma}^y_2$ at the
crossing point of paths $P_h$ and $P_v$. Thus,
$|\Psi_f\rangle=U_h^{-1}U_v^{-1}U_hU_v|g\rangle
=(\tilde{\sigma}^z_2)^{-\frac{1}{2}}[i(\tilde{\sigma}^z_2)^{-\frac{1}{2}}]|g\rangle
=i|\widetilde{Z}_2\rangle$, similar to the case with an $e$ vortex
looping around a superposition state of an $m$ vortex and the
vacuum~\cite{Sarma07}. However, in the case without braiding [see
Fig.~\ref{fig2}(b)],
$|\Psi_f\rangle=U_h^{-1}U_v^{-1}U_hU_v|g\rangle=|g\rangle$.
Therefore, the braiding of topological excitations can be
distinguished by verifying if an excited state
$|\widetilde{Z}_2\rangle$ occurs at the crossing point of paths
$P_h$ and $P_v$.

{\it Discussion and conclusion.}{\bf---}When the magnetic flux in
the SQUID loop of each charge qubit is set to $\Phi_e=\Phi_0/2$,
then $h_x=0$ and $J_y=0$. Also, we assume that $E_c\gg J_x,J_z$.
When $\Phi_e=\Phi_0/2$, one can shift the gate voltage, at the $i$th
lattice point, far away from the usual working point $n_g\sim
\frac{1}{2}$ of the Kitaev lattice for a period of time $\tau$. This
yields a local $z$-type rotation on the $i$th qubit:
$R_i^z(\theta)=\exp[-i(h_z\tau/\hbar)\sigma_i^z]\equiv\exp(-i\theta\sigma_i^z/2)$.
When $\theta\equiv 2h_z\tau/\hbar=\pi$ (where $n_g>\frac{1}{2}$),
$R_i^z(\pi)=-i\sigma_i^z$, so the $\sigma_i^z$ operation on the
$i$th qubit is given by $\sigma_i^z=e^{i\pi/2}R_i^z(\pi)$, while
half of the rotation is
$(\sigma_i^z)^{\frac{1}{2}}=e^{i\pi/4}R_i^z(\pi/2)$. The
corresponding inverse rotations can be achieved by shifting the gate
voltage to $n_g<\frac{1}{2}$. Similarly, when $n_g=\frac{1}{2}$ and
$\Phi_e=0$, one has $h_z=0$, $h_x=-E_J$, and $J_x=0$. Here we now
assume that $E_J\gg |J_y|,J_z$. When $n_g=\frac{1}{2}$, we also
switch off the flux in the SQUID loop of the $i$th qubit for a time
$\tau$ (the working point of this Kitaev lattice is usually at
$0<\Phi_e<\Phi_0/2$). This produces a local $x$-type rotation on the
$i$th qubit: $R_i^x(\theta)=\exp[i(\delta
E_J\tau/\hbar)\sigma_i^x]\equiv\exp(i\theta\sigma_i^x/2)$, where
$\delta E_J=E_J-\frac{1}{2}E_J(\Phi_e)$. The $\sigma_i^x$ rotation
on the $i$th qubit is $\sigma_i^x=e^{-i\pi/2}R_i^x(\pi)$, where
$2\delta E_J\tau/\hbar=\pi$. With both $\sigma_i^z$ and $\sigma_i^x$
rotations available for the $i$th qubit, the $\sigma_i^y$ rotation
is given by $\sigma_i^y=e^{-i\pi/2}\sigma_i^z\sigma_i^x$. Therefore,
one can construct the operations $\tilde{\sigma}_i^z$ and
$\tilde{\sigma}_i^z$ for generating topological excitations by using
the single-qubit rotations $\sigma_i^z$ and $\sigma_i^x$.

In order to obtain accurate $z$- and $x$-type single-qubit
rotations, we assume that $E_c$ and $E_J$ are much larger than the
inter-qubit coupling. Actually, this strict condition can be
loosened for realistic systems. As shown in \cite{Wei}, accurate
effective single-qubit rotations can still be achieved using
techniques from nuclear magnetic resonance when the inter-qubit
coupling is small (instead of much smaller than $E_c$ and $E_J$). In
the vortex- and bond-state cases studied here, to reveal the
braiding statistics of topological excitations, one should verify
the occurrence of the excited state $|\widetilde{Z}_2\rangle$ at the
crossing point of paths $P_h$ and $P_v$. This needs to distinguish
$|\widetilde{Z}_2\rangle$ from the ground state. Here the difference
between $|\widetilde{Z}_2\rangle$ and the ground state is the phase
flip, induced by the single-qubit rotation $\sigma_i^z$, on the
charge state of the $z$-bond at the crossing point. This phase flip
of the charge state could be measured, e.g., using state
tomography~\cite{Liu}.

In the region $J'_z\geq J'_x, |J_y|$,
it is shown~\cite{Vidal} that when acting on the ground state, the
spin-pair operations $\tilde{\sigma}^z_i$ and $\tilde{\sigma}^y_i$
generally create both vortex states (Abelian anyons) and fermionic
excitations. However, the weight of the anyons is dominant in the
low-energy subspace if the number $m$ of the spin-pair operations
for braiding anyons is not large and $J'_z\gg J'_x, |J_y|$. Let
$J'_x= |J_y|=J$. For example, when $m=25$, $\gamma\equiv J/{J'}_z$
can be $\gamma\sim 0.2$~\cite{Vidal}.

In conclusion, we have proposed an approach to realize the Kitaev
model on a honeycomb lattice using superconducting quantum circuits.
Two particular cases are studied to demonstrate the topological
states and the braiding statistics. Our approach provides an
experimentally realizable many-body system for demonstrating exotic
properties of these topological phases.

F.N. was supported in part by the NSA, LPS, ARO, and the NSF Grant
No.~EIA-0130383. J.Q.Y. and X.F.S. were supported by the ``973"
Program Grant Nos. 2009CB929300 and 2006CB921205, and the NSFC Grant
Nos. 10625416 and 10534060.


\end{document}